\documentclass{article}

\usepackage[a4paper, margin=2.5cm]{geometry}
\usepackage{graphicx}
\usepackage[numbers,sort&compress,square]{natbib}
\usepackage{booktabs} 
\usepackage{amsmath}
\usepackage{mathtools}

\DeclareMathOperator\erfcx{erfcx}
\DeclareMathOperator{\sgn}{sgn}
\usepackage{authblk}

\usepackage{tikz}
\usetikzlibrary{decorations.pathreplacing,angles,quotes}
\usepackage{xifthen}
\usetikzlibrary{arrows}
\usepackage{pgfplots} 
\pgfplotsset{compat=newest}
\pgfplotsset{plot coordinates/math parser=false}
\newlength\figureheight
\newlength\figurewidth 
\usetikzlibrary{plotmarks}

\begin{document}

%\title{Study of aerodynamic flutter instability using the actuator line method and the implementation of a wind energy harvesting mechanism}

\title{Study of flutter instability using the actuator line method for wind energy harvesting devices}

\author{Vitor G. Kleine$^{1}$ and Matias Herrera$^{2}$}

\affil{$^1$Division of Aeronautical and Aerospace Engineering, Instituto Tecnológico de Aeronáutica, São José dos Campos, Brasil}
\affil{$^2$Centro Tecnologico Aeroespacial, Universidad Nacional de La Plata, La Plata, Argentina.}

\date{}

\maketitle

%\ead{vitor.kleine@gp.ita.br}

%\author{Vitor G. Kleine$^{1}$ and Matias Herrera$^{2}$}

%\address{$^1$Division of Aeronautical and Aerospace Engineering, Instituto Tecnológico de Aeronáutica, São José dos Campos, Brasil}
%\address{$^2$Centro Tecnologico Aeroespacial, Universidad Nacional de La Plata, La Plata, Argentina.}

%\ead{vitor.kleine@gp.ita.br}

\begin{abstract}
 The suitability of the actuator line method (ALM) to predict flutter instability is theoretically studied by employing a two-dimensional linear model of the ALM undergoing harmonic motion. Three different analytical models of the ALM, including or not the non-circulatory and pitch-rate terms, are compared to Theodorsen’s theory. First, classical methods using Theodorsen's function are employed to calculate reference values of flutter velocity and frequency. Then, the theoretical response of the ALM is predicted by replacing Theodorsen's function in the lift and aerodynamic pitching moment models with the corresponding complex function that relates the lift calculated by an unsteady ALM and the quasi-steady lift in harmonic motion. This method is applied to an airfoil typical section and to an energy harvesting device based on aeroelastic vibrations of an airfoil. From the results, it is possible to conclude that the classical ALM does not accurately predict flutter. However, we show that an ALM that considers the pitch-rate and non-circulatory terms has the capability to reproduce the results of classical methods if the ratio between ALM smearing parameter and chord is carefully chosen. These results can guide aeroelastic simulations of energy harvesting devices, large horizontal-axis wind turbines and fixed-wing aircraft.
\end{abstract}

\vspace{1em}

\section{Introduction}

The actuator line method has been proposed by~\cite{sorensen2002numerical} to simulate wakes of wind turbines. By modeling the blades using body forces, it greatly reduces the computational costs of aerodynamic simulations compared to blade-resolved computational fluid dynamics (CFD). Another advantage of the ALM is reducing the complexity of the mesh. By representing the airfoil by body forces, applied to internal points or volumes of the computational domain, the user does not need to generate a mesh that conforms to the geometry of the blades. These two advantages of the method (reduced computational cost and reduced complexity of the mesh) make it very suitable for aeroelastic studies. Therefore, ALM has been used as the aerodynamic model for aeroelastic studies since its first applications~\cite{mikkelsen2003actuator,sorensen2015simulation}.

Despite its widespread application for aeroelastic studies, which are highly dependent on unsteady aerodynamic loads, only recently~\cite{taschner2025unsteady,alva2026applicability}, the unsteady effects of the ALM have been studied in a systematic approach and from a theoretical perspective. Taschner et al.~\cite{taschner2025unsteady} and Alva et al.~\cite{alva2026applicability} showed that, for low frequencies, the ALM results tend to a quasi-steady approach, as expected. However, Alva et al.~\cite{alva2026applicability} also showed that, to accurately reproduce results of the classical Theodorsen's theory~\cite{theodorsen1935general}, the user should carefully choose the ratio between smearing parameter and airfoil cord ($\varepsilon/c$) and include the pitch-rate term in the calculation of the lift force. Failing to include the pitch-rate term or an inadequate choice of smearing parameter would lead to errors in the unsteady loads for higher frequencies.

The relatively low level of flexibility of the blades and low frequencies involved in the most relevant aeroelastic mechanisms of traditional horizontal-axis wind turbines (HAWT) might have obscured the effects identified in~\cite{alva2026applicability}, for validations of aeroelastic effects such as~\cite{trigaux2024investigation,hodgson2022validation}. However, as the diameter of the blades increases, flexibility and aeroelastic problems become more important. Therefore, it is important to understand the limitations of the ALM to accurately predict fluid-structure interactions at higher frequencies. One of the aeroelastic effects that had a lower relevance is flutter. As can be observed by the analysis of Chetan et al.~\cite{chetan2022flutter}, flutter speeds were predicted to be well above operational speeds for HAWTs with blade length around 60 m. However, flutter in HAWT is becoming a more relevant topic of research~\cite{hayat2016flutter,chetan2022flutter,zhuang2024study} as the flexibility of blades increases.

Besides large horizontal-axis wind turbines, another wind energy technology that is highly dependent on fluid-structure interactions is flutter-based energy harvesting devices~\cite{erturk2010energy,li2022recent}. In this concept, the energy that is transferred from the flow to the structure through an unstable flutter mode is harvested to generate electricity. For this type of energy harvesting device, accurate computation of aeroelastic effects for moderately high reduced frequencies is paramount to the viability of the technology. For this reason, we chose a configuration of a 2-d airfoil coupled to a piezoelectric device, described in~\cite{erturk2010energy}, to investigate the suitability of ALM to predict flutter instability and accurately compute unsteady aerodynamic effects in moderately-high reduced frequencies.

In the current work, we study the aerodynamic flutter instability theoretically. First, the airfoil motion equations are analyzed using methods found in the literature~\cite{fung2008introduction,erturk2010energy} to calculate flutter velocity and frequency, using Theodorsen's functions~\cite{theodorsen1935general}. Then, the theoretical response of the ALM is predicted by replacing the classical Theodorsen's function in the lift and aerodynamic moment calculations with the complex function developed in~\cite{alva2026applicability}, which relates the induced velocities at the actuator lines to the circulation of an airfoil in harmonic motion. This complex function considers the effect of the ALM Gaussian smearing parameter $\varepsilon$, which plays a crucial role in the setup of numerical simulations.

%1 - Our goal is to provide useful guidelines on how to perform CFD simulations using the ALM that could predict flutter as accurately as Theodorsen's theory. However, it should be noted that the present work is a theoretical study and the execution of CFD simulations is left for future studies.

Our goal is to provide useful guidelines on how to perform CFD simulations using the ALM that could predict flutter as accurately as Theodorsen's theory. However, it should be noted that the present work is a theoretical study and the execution of CFD simulations is left for future studies. The behavior of the ALM is predicted by variants of the linear model proposed by~\cite{alva2026applicability}, using a MATLAB script, which has been shown to predict the results of ALM simulations very well~\cite{alva2026applicability}, at least for the low amplitudes which are expected to happen at the onset of flutter instabilities.

We analyze three different models of ALM and the influence of $\varepsilon$ in each one. In the first model, we consider the classical ALM, which calculates the lift force based on the local effective angle of attack. In the second model, we add the pitch-rate ($\dot{\alpha}$) term, which was shown by~\cite{alva2026applicability} to be fundamental for accurate computation of unsteady loads. In the third model, we add the non-circulatory terms predicted by Theodorsen's theory~\cite{theodorsen1935general}.

Finally, a wind energy harvesting mechanism based on aeroelastic vibrations was implemented following the 2-d model proposed by Erturk et al.~\cite{erturk2010energy}. We compare the results of this energy harvesting device with the results predicted by the ALM model.

In summary, our goal is to understand which terms of the lift and aerodynamic pitching moment should be included in the ALM computation, beyond the classical ALM, and which choice of $\varepsilon/c$ could lead to acceptable flutter predictions. As far as we are aware, this is the first study to analyze the suitability of ALM to predict flutter. The knowledge obtained in this study might be employed not only for vibration-based energy harvesting devices but also for HAWT. Additionally, since ALM has been shown to also be capable of accurately simulating lifting surfaces of airplanes in steady flow~\cite{kleine2023simulating}, its application for aeroelastic behavior of fixed-wing aircraft depends on careful validation of flutter predictions.

\section{Methodology}
\subsection{Equations of motion and structural model} \label{sec:structuralmodel}
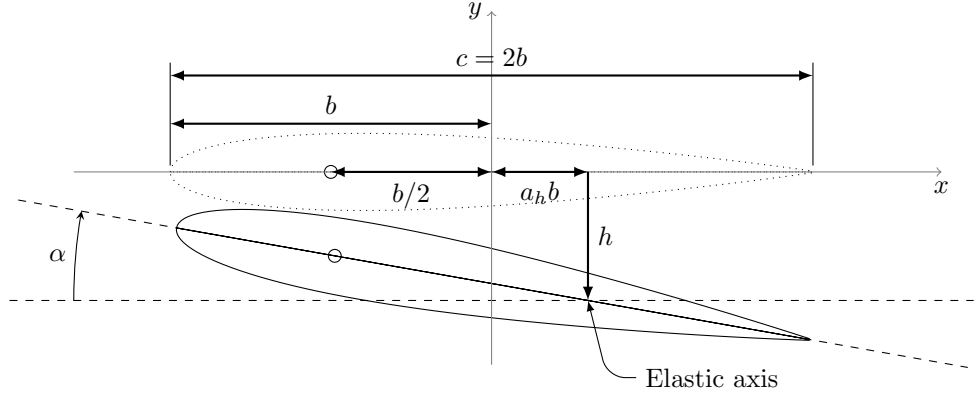
\begin{figure}
    \centering
    \begin{tikzpicture}
  \def\c{8.5}
  \def\r{0.80*\c}
  \def\ah{0.15}
  \def\a{\ah*\c}
  \def\hh{-0.2}
  \def\h{\hh*\c}
  % Some profiles look better when using plot[smooth]
  \draw[scale=\c,shift={(-0.5,\hh)},rotate around={-10:(0.5+\ah,0)}] plot file{Figures/NACA0012.tex} -- cycle;
  \draw[scale=\c,shift={(-0.5,0)},rotate around={-0:(0.5+\ah,\hh)},dotted] plot file{Figures/NACA0012.tex} -- cycle;
  \draw[shift={(\a,\h)},rotate=-10] (-0.25*\c-\a,0) circle [radius=0.01*\c];
  \draw[shift={(\a,0)},rotate=-0] (-0.25*\c-\a,0) circle [radius=0.01*\c];
  \draw[shift={(\a,\h)},rotate=-10,dashed] (-0.9*\c,0)--(0.6*\c,0);
  \draw[shift={(\a,\h)},dashed] (-0.9*\c,0)--(0.6*\c,0);
  \draw[->,gray] (-0.65*\c,0)--(0.70*\c,0) node[below,black]{$x$};
  \draw[->,gray] (0,-0.3*\c)--(0,0.25*\c) node[left,black]{$y$};
  %\draw[->] (0,0)--(0.7*\c,0) node[below]{$u_z$};
  %\draw[-,very thick] (-0.5*\c,0)--(0.5*\c,0);
  \draw[-latex,thick] (\a,0)--(\a,\h) node[right,midway]{$h$};
  \draw[latex-latex,thick] (0,0)--(\a,0) node[below,midway]{$a_h b$};
  \draw[latex-latex,thick] (-0.25*\c,0)--(0,0) node[below,midway]{$b/2$};
  \draw[latex-latex,thick] (-0.5*\c,0.15*\c)--(0.5*\c,0.15*\c) node[above,midway]{$c = 2b$};
  \draw[latex-latex,thick] (-0.5*\c,0.075*\c)--(0,0.075*\c) node[above,midway]{$b$};
  \draw[-] (-0.5*\c,0.01*\c)--(-0.5*\c,0.17*\c);
  \draw[-] (0.5*\c,0.01*\c)--(0.5*\c,0.17*\c);
  \draw[latex-,rounded corners=0.8*\c] (\a,\h) to (1.2*\a,1.6*\h) to (1.5*\a,1.6*\h) node[right]{Elastic axis};
  %\draw[->] (0,0)--(0.7*\c,0.2*\c) node[right]{$u_{r}$};
  %\draw[dashed] (0,0)--(-0.7*0.75*\c,-0.2*0.75*\c);
  %\draw[->,very thick] (0,0)--(0.7*0.12*\c,0.2*0.12*\c) node[above]{$F_{d}$};
  %\draw[->,very thick] (0,0)--(-0.2*0.7*\c,0.7*0.7*\c) node[left]{$F_{l}$};
  \draw[-stealth] (\a,\h) +(180:\r) arc(180:170:\r) node[left,midway]{$\alpha$};
  %\draw[-stealth] (0,0) +(195.945:1.4*\r) arc(195.945:170:1.4*\r) node[left,midway]{$\alpha$};
\end{tikzpicture}
    \caption{Notations used for the geometric parameters of the two-dimensional airfoil. Dotted line indicates mean position of airfoil and full line indicates displaced airfoil. Circle indicates the quarter-of-chord, where the circulatory lift is assumed to act.}
    \label{fig:airfoilnotation}
\end{figure}

First, we have analyzed the classical 2-d two-degree-of-freedom equations of motion of an airfoil (typical section)~\cite{fung2008introduction}:
\begin{equation}
m\ddot{h} + S\ddot{\alpha} + h k_h = -L,
\end{equation}
\begin{equation}
S\ddot{h} + I_{\alpha}\ddot{\alpha} + \alpha k_\alpha = M,
\end{equation}
where $h$ is the plunge displacement (translation), $\alpha$ is the pitch displacement (rotation), $m$ is the airfoil mass per unit span, $S$ is the static mass moment per unit span, $I_{\alpha}$ is the moment of inertia per unit span, $k_h$ is the stiffness per unit span in the plunge DOF, $k_\alpha$ is the stiffness per unit span in the pitch DOF, $L$ is the aerodynamic lift per unit span and $M$ is the aerodynamic pitching moment per unit span. The over-dot represents differentiation with respect to time. The notation for the airfoil geometric parameters is shown in figure~\ref{fig:airfoilnotation}.

\subsection{Wind energy harvesting mechanism} \label{sec:harvestingmodel}
In addition to the cases analyzing flutter of a typical section, we have implemented a piezoelectric mechanism capable of extracting energy from the flow, following the implementation of the model carried out by Erturk et al.~\cite{erturk2010energy}. This implementation consists of a third differential equation simultaneously solved with the airfoil’s equations of motion to consider the piezoaeroelastic coupling. The equations of motion of the airfoil are modified to
\begin{equation}
(m+m_f)\ddot{h} + S\ddot{\alpha} + h k_h + d_h \dot{h} - \theta \frac{v}{l}  = -L,
\end{equation}
\begin{equation}
S\ddot{h} + I_{\alpha}\ddot{\alpha} + \alpha k_\alpha + d_{\alpha} \dot{\alpha} = M,
\end{equation}
where $m_f$ accounts for a fixture mass per unit span in the experiments connecting the airfoil to the plunge springs, $d_h$ and $d_{\alpha}$ are structural damping coefficients, $v$ is the electric voltage across the resistive load, $l$ is the span length and $\theta$ is the electromechanical coupling term. The equation for the piezoelectric component is
\begin{equation}
C_p^{eq} \dot{\psi} + \frac{v}{R_l} + \theta\dot{h} = 0 ,
\end{equation}
where $C_p^{eq}$ is the equivalent capacitance of the piezoelectric material and $R_l$ is the electrical resistance. The reader is referred to Ref.~\cite{erturk2010energy} for schematics and details on this model.

%Rev 1: 2 - The reader is referred to ref.~\cite{erturk2010energy} for details about this model.

\subsection{Aerodynamic models} \label{sec:aerodynamicmodel}

\subsubsection{Classical Theodorsen's model}
In the classical Theodorsen's model~\cite{theodorsen1935general}, lift and moment can be written as:
\begin{equation}
L = L_{NC} + C(k) L_{QS}
\label{eq:classicalL}
\end{equation}
\begin{equation}
M = M_{NC} + C(k) M_{QS}
\label{eq:classicalM}
\end{equation}
where $L_{NC}$ and $M_{NC}$ are the non-circulatory terms, $L_{QS}$ and $M_{QS}$ are the quasi-steady circulatory terms, $C(k)$ is Theodorsen's function and $k$ is the reduced frequency. They are defined as~\cite{bisplinghoff1996aeroelasticity,fung2008introduction}\footnote{Some typos found in previous versions of this paper were corrected.}:
\begin{equation}
    L_{NC} = \pi \rho b^2 (\ddot{h}+U \dot{\alpha}-b a_h \ddot{\alpha})
\end{equation}
\begin{equation}
    M_{NC} = \pi \rho b^2 (b a_h \ddot{h}-Ub(1/2-a_h) \dot{\alpha} -b^2 (1/8+a_h^2) \ddot{\alpha})
\end{equation}
\begin{equation}
    L_{QS} = 2 \pi \rho U b (\dot{h}+U \alpha + b (1/2-a_h) \dot{\alpha})
    \label{eq:Lqs}
\end{equation}
\begin{equation}
    M_{QS} = b (a_h+1/2) L_{QS}
\end{equation}
\begin{equation}
    C(k) = \frac{H_1^{(2)}(k)}{H_1^{(2)}(k)+i H_0^{(2)}(k)}
\end{equation}
\begin{equation}
    k = \frac{\Omega b}{U}
\end{equation}
where $\rho$ is the density, $b=c/2$ is the semichord, $U$ is the freestrem velocity, $\Omega$ is the angular frequency of the harmonic motion and $a_h$ is the non-dimensional position of the axis of rotation. Due to the use of Theodorsen's theory as a reference case, only incompressible cases under the inviscid approximation are considered.

\subsubsection{Linear models for the unsteady actuator line method}

In the ALM, the forces are distributed in a volume in the numerical domain. To avoid numerical instabilities, the forces are spread using a length scale that should be a multiple of the grid spacing. For Gaussian functions, the smearing parameter $\varepsilon$ is usually taken to be at least 2 times the grid spacing~\cite{troldborg2009actuator} to avoid numerical instabilities, but values of around 4 times the grid spacing provide better computation of the angle of attack and resolution of vortices~\cite{shives2013mesh,kleine2023non}. For these reasons, the choice of $\varepsilon$ may be primarily dictated by computational costs rather than physical considerations. Anyhow, being a theoretical analysis, this study is not limited by constraints related to grid requirements or computational costs. As shown by~\cite{kleine2022stability,taschner2025unsteady,alva2026applicability}, the smearing parameter affects the distribution of vorticity and induced velocities in the unsteady case.

A linear model for the two-dimensional (2-d) actuator line method undergoing harmonic motion was developed by~\cite{alva2026applicability} (see also~\cite{taschner2025unsteady} for an independent derivation of the model). In this linear model, based on the vorticity generated by unsteady motion, the unsteady component of the lift always generates vorticity and sheds it to the wake~\cite{kleine2022stability}. Therefore, the lift force corresponds only to the circulatory term, and the non-circulatory terms of the lift force are absent, which can also be understood as a consequence of the absence of a physical airfoil in the numerical simulation.

The ALM model considered in~\cite{alva2026applicability} applies a 2-d isotropic Gaussian with smearing parameter $\varepsilon$ to spread the lift force in the CFD domain, and samples the velocity at the actuation point. Different shapes of the smearing function or sampling points might lead to different models. For conciseness, only the main results and formulas of~\cite{alva2026applicability} are shown here, and the reader is referred to the original article for the complete description and details of the linear model and configuration of the ALM.

As shown in~\cite{alva2026applicability}, the behavior of the ALM can be well predicted by a linear model that employs a complex function $C_{\varepsilon}(\varepsilon/c,k)$ to relate the lift force to the quasi-steady lift in harmonic motion
\begin{equation}
L = C_{\varepsilon}(\varepsilon/c,k) L_{QS} .
\end{equation}
The complex function $C_{\varepsilon}(\varepsilon/c,k)$ is the ALM-equivalent to Theodorsen's function, and it is defined based on the following relation:
\begin{equation}
  C_{\varepsilon}(\varepsilon/c,k) = \frac{1}{1 - \frac{a_0 c}{2 \varepsilon} \kappa}
\end{equation}
where $a_0$ is the slope of $C_l$ vs $\alpha$, c is the chord and $\kappa(\varepsilon/c,k)$ is the complex function defined in~\cite{alva2026applicability}:
\begin{equation}
  \label{eq:kappa}
  \kappa(k_{\varepsilon}) = - \frac{i 2k_{\varepsilon} \exp(-k_{\varepsilon}^2) }{2 \sqrt{\pi}} \int\limits_{-\infty}^{+\infty} H_{\varepsilon}\left(z-i k_{\varepsilon} \right) \exp(-i 2 k_{\varepsilon} z) \sgn(z) \erfcx(|z|) dz .
\end{equation}
where $k_{\varepsilon}=k/(\varepsilon/c) = \Omega \varepsilon/(2 U)$ is the reduced frequency based on the smearing parameter.

%4 - In order to be compatible with Theodorsen's theory, the lift coefficient slope is taken as $a_0 = 2\pi$, from thin airfoil theory, and drag is neglected.

In order to be compatible with Theodorsen's theory, the lift coefficient slope is taken as $a_0 = 2\pi$, from thin airfoil theory, and drag is neglected.

Since the center of the bound vortex generated in the ALM is the center of the Gaussian~\cite{forsythe2015coupled,martinez2017optimal}, even for unsteady lift~\cite{kleine2022stability}, we consider that the lift force acts at the quarter-of-chord of the airfoil, similarly to the classical interpretation of the lifting line method~\cite{katz1991low}. Therefore, the circulatory aerodynamic moment is directly calculated by the ALM from:
\begin{equation}
    M = b (a+1/2) L = C_{\varepsilon}(\varepsilon/c,k) M_{QS} .
\end{equation}

As shown in~\cite{alva2026applicability}, the behavior of the ALM simulations can be well predicted by this linear model. One of the main conclusions of that work is that, to accurately represent unsteady effects, the pitch-rate term should be considered in equation~\ref{eq:Lqs}. However, the classical ALM does not include the pitch-rate term, considering only the effective angle of attack, adopting a formulation similar to a steady state approximation, which, for this case, can be written as:
\begin{equation}
    L_{SS} = 2 \pi \rho U b (\dot{h}+U \alpha) .
    \label{eq:Lss}
\end{equation}
Alva et al.\cite{alva2026applicability} also showed that the classical ALM can be modeled by
\begin{equation}
L = C_{\varepsilon}(\varepsilon/c,k) L_{SS}
\end{equation}
even though the agreement with Theodorsen's theory is poor.

Both the classical ALM and the modified ALM with the pitch-rate term do not consider non-circulatory terms. In order to evaluate the effect of this simplification for flutter instability and wind energy harvesting devices, we explicitly include the non-circulatory terms in our model, modeling the lift force and aerodynamic moment as
\begin{equation}
    L = L_{NC} + C_{\varepsilon}(\varepsilon/c,k) L_{QS}
\end{equation}
\begin{equation}
    M = M_{NC} + C_{\varepsilon}(\varepsilon/c,k) M_{QS} .
\end{equation}
In an ALM coupled with a CFD solver, we believe that these non-circulatory terms should not be included as part of the lift coefficient used to calculate body forces for the flow solver; otherwise, they would shed vorticity onto the wake. The non-circulatory terms should be added after the calculation of the ALM (circulatory) lift and used as input only for the structural dynamics.

\subsection{Flutter calculations} \label{sec:flutter}
In order to calculate the critical frequency and flutter velocity of the classical 2-d equation of motion of an airfoil, the method described in Chapter 6 of~\cite{fung2008introduction} was implemented in a MATLAB script. By substituting the lift and aerodynamic moment into the equations of motion, detailed in section~\ref{sec:structuralmodel}, and setting the determinant to zero, the critical frequency and flutter velocity are obtained for a specific airfoil geometry and set of parameters.

For the equations of the energy harvesting device, described in section~\ref{sec:harvestingmodel}, an iterative solution procedure is required to calculate the critical frequency and flutter velocity, as described by~\cite{erturk2010energy}.

In this work, the critical frequency and flutter velocity obtained using the three different levels of complexity of the ALM are compared to those derived from the classical Theodorsen's theory. To implement the ALM linear model instead of Theodorsen's model in the MATLAB script, the function $C_{\varepsilon}(\varepsilon/c,k)$ is used instead of $C(k)$, and some of the terms of the lift and moment are turned off, according to the models described in section~\ref{sec:aerodynamicmodel}, which are summarized below:
\begin{itemize}
    \item{\textbf{Reference case:}} By using Theodorsen’s function $C(k)$, considering the pitch-rate (term with $\dot{\alpha}$ in $L_{QS}$) and non-circulatory terms, the classical results of flutter are obtained, which are taken as the reference case. The lift and aerodynamic pitching moment are calculated from
    \begin{equation}
        L = L_{NC} + C(k) L_{QS}
    \end{equation}
    \begin{equation}
        M = M_{NC} + C(k) M_{QS} .
    \end{equation}
    \item{\textbf{Classical ALM}} It considers only the circulatory terms, without the pitch-rate term
    \begin{equation}
        L = C_{\varepsilon}(\varepsilon/c,k) L_{SS}
    \end{equation}
    \begin{equation}
        M = C_{\varepsilon}(\varepsilon/c,k) M_{SS} .
    \end{equation}
    \item{\textbf{ALM with pitch-rate term}} This corresponds to the ALM with pitch-rate term, following~\cite{alva2026applicability}, without modeling the non-circulatory terms
    \begin{equation}
        L = C_{\varepsilon}(\varepsilon/c,k) L_{QS}
    \end{equation}
    \begin{equation}
        M = C_{\varepsilon}(\varepsilon/c,k) M_{QS} .
    \end{equation}
    \item{\textbf{ALM with pitch-rate term and non-circulatory terms}} This is the most complete model of ALM for unsteady effects, including $C_{\varepsilon}(\varepsilon/c,k)$, the pitch-rate term ($\dot{\alpha}$) and both non-circulatory terms
    \begin{equation}
        L = L_{NC} + C_{\varepsilon}(\varepsilon/c,k) L_{QS}
    \end{equation}
    \begin{equation}
        M = M_{NC} + C_{\varepsilon}(\varepsilon/c,k) M_{QS} .
    \end{equation}
\end{itemize}

It should be noted that all the results were obtained using the linear model of the ALM implemented in a MATLAB script. We did not run any CFD simulations in the present work. Future CFD simulations are required to confirm the results presented here.

\section{Results}
\subsection{Effects of non-circulatory terms and pitch rate on the flutter instability} \label{sec:airfoilflutter}
First, we study the flutter instability of a two-dimensional airfoil (typical section). The results of a reference example involving flutter of a two-dimensional airfoil described in~\cite{fung2008introduction} are reproduced. The parameters for this case are shown in table~\ref{tab:parametersflutter}.
\begin{table}[b]
    \centering
    \caption{Parameters for flutter of a two-dimensional airfoil, from~\cite{fung2008introduction}. The physical interpretation of these terms can be found in~\cite{fung2008introduction}.}
    \begin{tabular}{c c c c c c c}
        \toprule
        $b = \frac{c}{2}$ & $\mu = \frac{m}{\pi \rho b^2}$ & $a_h$ & $x_{\alpha} = \frac{S}{m b}$ & $r_{\alpha}^2 = \frac{I_{\alpha}}{m b^2}$ & $\omega_{\alpha} = \sqrt{\frac{k_{\alpha}}{I_{\alpha}}}$ & $\omega_{h} = \sqrt{\frac{k_{h}}{m}}$ \\ \hline
        0.127 m & 76 & -0.15 & 0.25 & 0.388 & 64.1 rad/s & 55.9 rad/s \\
        \bottomrule
    \end{tabular}
    \label{tab:parametersflutter}
\end{table}

\begin{figure}[tb]
  \centering
\definecolor{mycolor1}{rgb}{0.0,0.2,0.9}%
\definecolor{mycolor2}{rgb}{0.9,0.2,0.0}%
\begin{tikzpicture}

\begin{axis}[%
width=4.5in,
height=2.5in,
at={(0.758in,0.481in)},
scale only axis,
xmin=0.2,
xmax=0.7,
xlabel style={font=\color{white!15!black}},
xlabel={$\varepsilon/c$},
every outer y axis line/.append style={mycolor2},
every y tick label/.append style={font=\color{mycolor2}},
every y tick/.append style={mycolor2},
ymin=0.399,
ymax=2.2,
ylabel style={font=\color{mycolor2}},
ylabel={$U/U^*$},
%axis background/.style={fill=white},
axis x line*=bottom,
axis y line*=right,
xmajorgrids,
ymajorgrids,
ytick distance=0.2,
legend style={legend cell align=left, align=left, draw=white!15!black}
]
\addplot [color=mycolor2, dashed, mark=o, mark options={solid, mycolor2}, forget plot]
  table[row sep=crcr]{%
0.25	0.399\\
0.33	0.459\\
0.4	0.503\\
0.5	0.557\\
0.6	0.603\\
0.7	0.64\\
};

\addplot [color=mycolor2, dashed, mark=x, mark options={solid, mycolor2}, mark size=3pt, forget plot]
  table[row sep=crcr]{%
0.25	0.795\\
0.33	0.858\\
0.4	0.899\\
0.5	0.938\\
0.6	0.959\\
0.7	0.944\\
};

\addplot [color=mycolor2, dashed, mark=square, mark options={solid, mycolor2}, forget plot]
  table[row sep=crcr]{%
0.25	1.01\\
0.33	1.006\\
0.4	0.995\\
0.5	0.963\\
0.6	0.897\\
};

\addplot [color=black, line width=1.0pt, forget plot]
  table[row sep=crcr]{%
0.2	1\\
0.7	1\\
};
\end{axis}

\begin{axis}[%
width=4.5in,
height=2.5in,
at={(0.758in,0.481in)},
scale only axis,
xmin=0.2,
xmax=0.7,
xlabel style={font=\color{white!15!black}},
xlabel={$\varepsilon/c$},
every outer y axis line/.append style={mycolor1},
every y tick label/.append style={font=\color{mycolor1}},
every y tick/.append style={mycolor1},
ymin=0.399,
ymax=2.2,
ylabel style={font=\color{mycolor1}},
ylabel={$k/k^*$},
%axis background/.style={fill=white},
axis x line*=bottom,
axis y line*=left,
%xmajorgrids,
%ymajorgrids,
ytick distance=0.2,
legend style={legend cell align=left, align=left, draw=white!15!black}
]
\addplot [color=mycolor1, mark=o, mark options={solid, mycolor1}]
  table[row sep=crcr]{%
0.25	2.115\\
0.33	1.847\\
0.4	1.689\\
0.5	1.533\\
0.6	1.425\\
0.7	1.349\\
};
\addlegendentry{Classical ALM}

\addplot [color=mycolor1, mark=x, mark options={solid, mycolor1}, mark size=3pt]
  table[row sep=crcr]{%
0.25	1.113\\
0.33	1.051\\
0.4	1.022\\
0.5	1.009\\
0.6	1.025\\
0.7	1.106\\
};
\addlegendentry{ALM with $\dot{\alpha}$ without NC}

\addplot [color=mycolor1, mark=square, mark options={solid, mycolor1}]
  table[row sep=crcr]{%
0.25	0.956\\
0.33	0.985\\
0.4	1.02\\
0.5	1.095\\
0.6	1.234\\
};
\addlegendentry{ALM with $\dot{\alpha}$ and NC}
\end{axis}

\end{tikzpicture}%
  \caption{Graphical representation of ratio of flutter frequencies (full lines) and flutter speeds (dashed lines), for the typical section. For perfect agreement with Theodorsen's theory, both ratios should be 1.}
  \label{fig:Flutter_results_airfoi}
\end{figure}
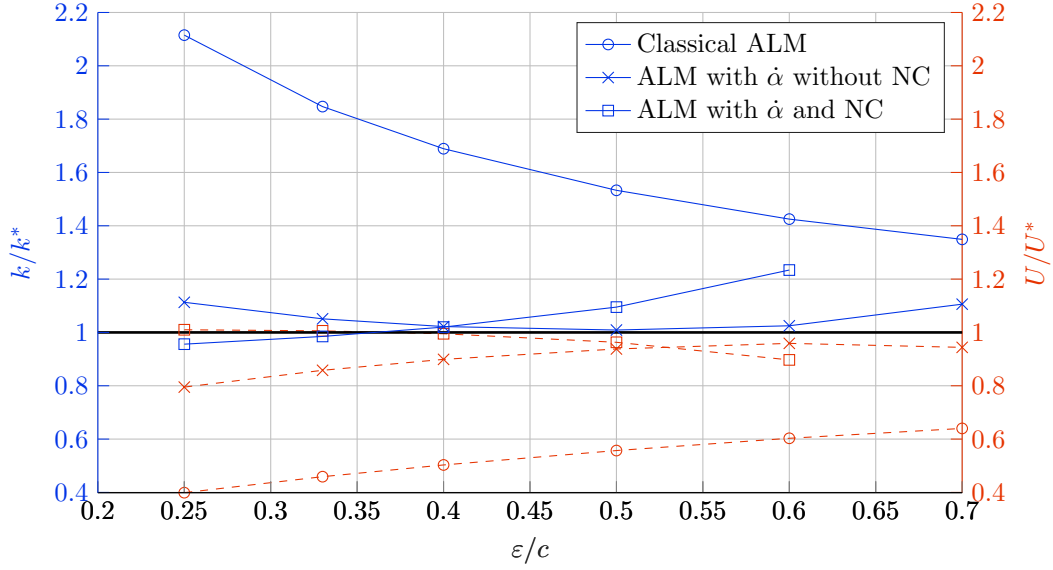

In this example, the reduced frequency $k^*$ is equal to 0.274 and the critical flutter velocity $U^*$ is 27.712 m/s, calculated using Theodorsen's theory. In table~\ref{tab:flutter_instability}, the results obtained for all the ALM models are shown, for different ratios of $\varepsilon/c$. These results are reproduced in graphical form in figure~\ref{fig:Flutter_results_airfoi}, for the most relevant range of $\varepsilon/c$. As indicated in table~\ref{tab:flutter_instability}, for some of the values of $\varepsilon/c$, no solution was found, which indicates that flutter instability would not occur in that ALM model.

\begin{table}[tb]
\centering
\caption{Effects of pitch-rate term ($\dot{\alpha}$) and non-circulatory terms (NC) on the flutter instability. Reference values calculated using Theodorsen's theory are $k^*=0.274$ and $U^*=27.71$ m/s.}
\label{tab:flutter_instability}
\begin{tabular}{c|cc|cc|cc}
\toprule
$\varepsilon/c$ & \multicolumn{2}{c|}{Classical ALM} & \multicolumn{2}{c|}{ALM with $\dot{\alpha}$ without NC} & \multicolumn{2}{c}{ALM with $\dot{\alpha}$ and NC} \\

 & $k/k^*$ & $U/U^*$ & $k/k^*$ & $U/U^*$ & $k/k^*$ & $U/U^*$ \\
\midrule
0.25 & 2.115 & 0.399 & 1.113 & 0.795 & 0.956 & 1.010 \\
0.33 & 1.847 & 0.459 & 1.051 & 0.858 & 0.985 & 1.006 \\
0.4  & 1.689 & 0.503 & 1.022 & 0.899 & 1.020 & 0.995 \\
0.5  & 1.533 & 0.557 & 1.009 & 0.938 & 1.095 & 0.963 \\
0.6  & 1.425 & 0.603 & 1.025 & 0.959 & 1.234 & 0.897  \\
0.7  & 1.349 & 0.640 & 1.106 & 0.944 & *     & *      \\
1    & 1.210 & 0.725 & *     & *     & *     & *      \\
2    & 1.102 & 0.835 & *     & *     & *     & *      \\
4    & 1.164 & 0.801 & *     & *     & *     & *      \\
\bottomrule
\multicolumn{7}{l}{\footnotesize $^*$The method did not find a solution for the determinant of the equations of motion.}
\end{tabular}
\end{table}

In the most complete ALM model, including both the pitch rate and non-circulatory terms, we obtained ratios $k/k^*$ and $U/U^*$ very close to 1 for $\varepsilon/c = 0.33$ and $\varepsilon/c = 0.4$. For greater or smaller $\varepsilon/c$, the flutter frequency and velocity deviate from the results predicted by the classical theory. These results agree very well with the results found by Alva et al.~\cite{alva2026applicability}, who found that the ALM has a great agreement with the circulatory terms of Theodorsen's theory for $\varepsilon/c$ between $1/3$ and $0.4$, for reduced frequencies lower than $0.5$.

The results from table~\ref{tab:flutter_instability} show that neglecting non-circulatory terms leads to larger errors. We obtained ratios $k/k^*$ close to 1 for $\varepsilon/c$ around $0.5$, but the errors in $U/U^*$ are around one order of magnitude higher than when the non-circulatory terms are considered. Nevertheless, an error of around 10\% might still be acceptable in some cases. A lower flutter velocity is being calculated by the ALM, which would lead to a conservative approach, from a design perspective. For the design of aeronautical wings or wind turbine blades, a conservative approach is desirable. However, for a flutter-based energy harvesting device, it might lead to an underprediction of power production. It should be noted that higher errors are expected for higher flutter frequencies when neglecting non-circulatory terms.

The classical ALM, however, does not satisfactorily predict flutter. This discrepancy can be attributed to the lack of the pitch-rate term, which has already been shown by~\cite{alva2026applicability} to be essential to accurately model the unsteady effects in ALM when there is pitch motion. Alva et al.~\cite{alva2026applicability} showed that the errors are especially pronounced in the phase difference, which is known to play an important role in flutter instability.

The particular cases where the solutions were not found are of special importance. Our interpretation is that flutter would not be triggered in ALM simulations, at least for small amplitudes of disturbances, despite being expected to occur, according to Theodorsen's theory. These cases highlight the importance of the smearing parameter. For larger values of the smearing parameter, the most complete ALM models do not predict flutter. Therefore, there is the risk of false negatives in flutter prediction using ALM simulations with inadequate models or values of $\varepsilon/c$.

\subsection{Flutter of wind energy harvesting device}\label{sec:windenergyharvesting}
The parameters of the piezoelectric energy harvesting device based on aeroelastic vibrations described in~\cite{erturk2010energy} are shown in table~\ref{tab:parametersharvesting}. Assuming an air density of $\rho=1.225$ kg/m$^3$ and using Theodorsen's model, our results match the critical reduced frequency calculated by~\cite{erturk2010energy}, $k^*$=0.425. A value of $U^*$=9.49 m/s was calculated for the critical flutter speed, compared to 9.56 m/s calculated by~\cite{erturk2010energy}. The reasons for the difference were not identified; however, given this small difference of 0.7\%, we considered it a good agreement.

\begin{table}[tb]
    \centering
    \caption{Parameters for the two-dimensional flutter-based energy harvesting device, consisting of a 2-d airfoil and a piezoelectric component, from~\cite{erturk2010energy}.}
    \begin{tabular}{c c c c c c c}
        \toprule
        $b = \frac{c}{2}$ & $\mu = \frac{m}{\pi \rho b^2}$ & $a_h$ & $x_{\alpha} = \frac{S}{m b}$ & $r_{\alpha}^2 = \frac{I_{\alpha}}{m b^2}$ & $\omega_{\alpha} = \sqrt{\frac{k_{\alpha}}{I_{\alpha}}}$ & $\omega_{h} = \sqrt{\frac{k_{h}}{m}}$ \\ \hline
        0.125 m & 29.6 & -0.5 & 0.26 & 0.254 & 15.4 rad/s & 51.28 rad/s\\
        \bottomrule
    \end{tabular}
    \begin{tabular}{c c c c c c c}
        \toprule
        $l$ & $\beta = \frac{m+m_f}{m}$ & $\gamma_h = \frac{\omega d_h}{k_h}$ & $\gamma_{\alpha} = \frac{\omega d_{\alpha}}{k_{\alpha}}$ & $\theta$ & $C_p^{eq}$ & $R_l$ \\ \hline
        0.5 m & 2.597 & 0.007 & 0.12 & 1.55 mN/V & 120 nF & 100 k$\Omega$ \\
        \bottomrule
    \end{tabular}
    \label{tab:parametersharvesting}
\end{table}

For this case, the relevance of the non-circulatory and pitch-rate terms is expected to be higher than the results from section~\ref{sec:airfoilflutter}, because they tend to grow with the increase of the reduced frequency. Hence, the classical ALM is left out of this analysis, due to the poor agreement shown in section~\ref{sec:airfoilflutter}, and only the values of $\varepsilon/c$ within the range of good results from table~\ref{tab:flutter_instability} are considered.

\begin{table}[tb]
\centering
\caption{Effects of non-circulatory terms (NC) on flutter frequency and velocity of the flutter-based energy harvesting device. Reference values calculated using Theodorsen's theory are $k^*=0.425$ and $U^*=9.49$ m/s.}
\label{tab:flutter_energyharvesting}
\begin{tabular}{c|cc|cc}
\toprule
$\varepsilon/c$ & \multicolumn{2}{c|}{ALM with $\dot{\alpha}$ without NC} & \multicolumn{2}{c}{ALM with $\dot{\alpha}$ and NC} \\
 & $k/k^*$ & $U/U^*$ & $k/k^*$ & $U/U^*$ \\
\midrule
0.25 & 0.391 & 1.518 & 0.875 & 1.128 \\
0.33 & 0.393 & 1.555 & 0.972 & 1.027 \\
0.4  & 0.395 & 1.583 & 1.054 & 0.954 \\
0.5  & 0.402 & 1.611 & 1.178 & 0.861 \\
0.6  & 0.414 & 1.626 & 1.308 & 0.781 \\
\bottomrule
%\multicolumn{5}{l}{\footnotesize $^*$The method did not find a solution for the determinant of the equations of motion.}
\end{tabular}
\end{table}

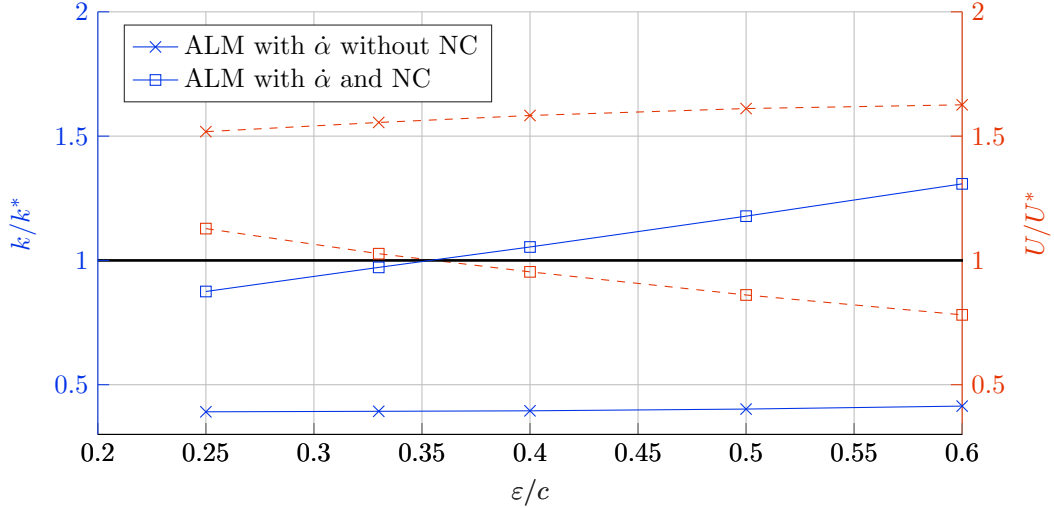
\begin{figure}[tb]
  \centering
\definecolor{mycolor1}{rgb}{0.0,0.2,0.9}%
\definecolor{mycolor2}{rgb}{0.9,0.2,0.0}%
\begin{tikzpicture}

\begin{axis}[%
width=4.5in,
height=2.2in,
at={(0.758in,0.481in)},
scale only axis,
xmin=0.2,
xmax=0.6,
xlabel style={font=\color{white!15!black}},
xlabel={$\varepsilon/c$},
every outer y axis line/.append style={mycolor2},
every y tick label/.append style={font=\color{mycolor2}},
every y tick/.append style={mycolor2},
ymin=0.3,
ymax=2.0,
ylabel style={font=\color{mycolor2}},
ylabel={$U/U^*$},
%axis background/.style={fill=white},
axis x line*=bottom,
axis y line*=right,
xmajorgrids,
ymajorgrids,
%legend pos=north west,
%legend style={legend cell align=left, align=left, draw=white!15!black}
]

\addplot [color=mycolor2, dashed, mark=x, mark options={solid, mycolor2}, mark size=3pt, forget plot]
  table[row sep=crcr]{%
0.25	1.518\\
0.33	1.555\\
0.4	1.583\\
0.5	1.611\\
0.6	1.626\\
};

\addplot [color=mycolor2, dashed, mark=square, mark options={solid, mycolor2}, forget plot]
  table[row sep=crcr]{%
0.25	1.128\\
0.33	1.027\\
0.4	0.954\\
0.5	0.861\\
0.6	0.781\\
};

\addplot [color=black, line width=1.0pt, forget plot]
  table[row sep=crcr]{%
0.2	1\\
0.7	1\\
};
\end{axis}

\begin{axis}[%
width=4.5in,
height=2.2in,
at={(0.758in,0.481in)},
scale only axis,
xmin=0.2,
xmax=0.6,
xlabel style={font=\color{white!15!black}},
xlabel={$\varepsilon/c$},
every outer y axis line/.append style={mycolor1},
every y tick label/.append style={font=\color{mycolor1}},
every y tick/.append style={mycolor1},
ymin=0.3,
ymax=2.0,
ylabel style={font=\color{mycolor1}},
ylabel={$k/k^*$},
%axis background/.style={fill=white},
axis x line*=bottom,
axis y line*=left,
%xmajorgrids,
%ymajorgrids,
legend pos=north west,
legend style={legend cell align=left, align=left, draw=white!15!black}
]

\addplot [color=mycolor1, mark=x, mark options={solid, mycolor1}, mark size=3pt]
  table[row sep=crcr]{%
0.25	0.391\\
0.33	0.393\\
0.4	0.395\\
0.5	0.402\\
0.6	0.414\\
};
\addlegendentry{ALM with $\dot{\alpha}$ without NC}

\addplot [color=mycolor1, mark=square, mark options={solid, mycolor1}]
  table[row sep=crcr]{%
0.25	0.875\\
0.33	0.972\\
0.4	1.054\\
0.5	1.178\\
0.6	1.308\\
};
\addlegendentry{ALM with $\dot{\alpha}$ and NC}
\end{axis}

\end{tikzpicture}%
  \caption{Graphical representation of ratio of flutter frequencies (full lines) and flutter speeds (dashed lines), for the flutter-based energy harvesting device. For perfect agreement with Theodorsen's theory, both ratios should be 1.}
  \label{fig:Flutter_results_energyharvesting}
\end{figure}

Results obtained using the ALM with and without non-circulatory terms are shown in table~\ref{tab:flutter_energyharvesting} and figure~\ref{fig:Flutter_results_energyharvesting}. These results confirm that this case is more challenging for an aerodynamic model. Neglecting the non-circulatory terms considerably changes the results of flutter calculations. This case also highlights the need to carefully choose the value of the ratio between the smearing parameter and the chord. For the most complete ALM model, very good agreement was found for $\varepsilon/c$ between 0.33 and 0.4, which is compatible with the best range of values identified in the literature for unsteady loads~\cite{alva2026applicability}. However, the results are very sensitive to the value of $\varepsilon/c$, which is to be expected, given the higher sensitivity to $\varepsilon/c$ of the unsteady results of the ALM for higher reduced frequencies~\cite{taschner2025unsteady,alva2026applicability}.

\section{Conclusions}
In the present work, we investigated, theoretically, the suitability of the actuator line method (ALM) to predict flutter instability. By using a two-dimensional linear model~\cite{alva2026applicability} that predicts the behavior of the ALM, the need to perform CFD simulations is avoided. The good agreement between the linear model and ALM simulations of oscillating airfoils, shown in past studies~\cite{taschner2025unsteady,alva2026applicability}, encourages the use of the results of this work to guide aeroelastic simulations. Nevertheless, future studies employing ALM in non-linear CFD simulations should be performed to confirm the predictions of the models.

From these theoretical results, we conclude that the classical ALM is not suitable for flutter analysis, due to the absence of the pitch-rate term. The results shown in this work indicate that, to accurately compute flutter velocity and frequency, an ALM implementation should:
\begin{itemize}
    \item Include the pitch-rate term in the calculation of the lift, as detailed in~\cite{alva2026applicability};
    \item Use an appropriate choice of smearing parameter, following the comparison shown in~\cite{alva2026applicability};
    \item Include a model of the non-circulatory terms.
\end{itemize}

The accuracy of the ALM in predicting the flutter frequency and velocity for energy harvesting devices based on aeroelastic vibrations of an airfoil was also studied. Using the most complete ALM model, our results indicate that the ALM can be used to study these types of devices, with an accuracy compatible with the classical Theodorsen's theory, when employing the three recommendations listed above.

In this work, a typical section and a flutter-based energy harvesting device were analyzed, but we believe the main conclusions are also valid for aeroelastic studies of large horizontal-axis wind turbines and aeronautical applications using the ALM.

%\section{Planned additions}
%In the final version of the paper, we will implement all three cases of study in the ALM code using Nek5000, and compare to the results of tables~\ref{tab:flutter_instability} and section ~\ref{sec:wind energy harvesting}, with the goal of understanding which terms should be kept in the model and the best choice of parameters to capture flutter and the unsteady response of an airfoil with and without the energy harvesting mechanism. Additionally, we will detail the methods and expand all sections.

%The main contribution of the present work is to provide guidelines on how to perform unsteady ALM simulations that can accurately capture the main aeroelastic effects.

%We are planning to add for full paper a more detailed study of the effects of non-circulatory terms and pitch ratio on the flutter instability. I need to add a more detailed explanation here.

%\subsection{Some comments on the formating of references}
%Follow carefully the instructions in “Basic guidelines for preparing a paper” + use numbered references + preferably include titles of the references + add DOI where possible. When using the latex template, consider using bibtex in combination with the iopart-num bibtex style \cite{wes-7-2491-2022,wes-9-2087-2024,wes-8-1071-2023}

%\ack
%Matias Herrera acknowledges the financial support received from CAPES (Coordenacao deAperfeicoamento de Pessoal de Nivel Superior) under a Grant in the program “Move la America” for academic research at Instituto Tecnologico de Aeronáutica (ITA) in Brazil.

\section*{Acknowledgements}
This study was partially funded by Finep and Embraer S.A., under the research project Advanced Studies in Flight Physics and Control – contract number 01.22.0552.00, and by Program Move La America of the Brazilian Federal Agency for Support and Evaluation of Graduate Education – CAPES.

\bibliographystyle{iopart-num}
\bibliography{Torque26_references}

\end{document}